\documentclass[aip, apl, reprint]{revtex4-2}
\usepackage{placeins}
\usepackage{amsmath,amsfonts,amssymb}
\usepackage{amsmath}
\usepackage{graphicx}
\usepackage[colorlinks=true, allcolors=blue]{hyperref}
\usepackage{siunitx}
\usepackage{changes}
\usepackage{float}
\usepackage{hyperref}
\graphicspath{{../Figures/PDF/}}
\definechangesauthor[name={Dmitry Bykov}, color=orange]{DB}

\begin{document}
\title{Hybrid electro-optical trap for experiments with levitated particles in vacuum}

\author{Dmitry S. Bykov}
\email[]{dmitry.bykov@uibk.ac.at}
\address{Institut f{\"u}r Experimentalphysik, Universit{\"a}t Innsbruck, Technikerstra\ss e 25, 6020 Innsbruck,
	Austria}
\author{Maximilian Meusburger}
\address{Institut f{\"u}r Experimentalphysik, Universit{\"a}t Innsbruck, Technikerstra\ss e 25, 6020 Innsbruck,
	Austria}
\author{Lorenzo Dania}
\address{Institut f{\"u}r Experimentalphysik, Universit{\"a}t Innsbruck, Technikerstra\ss e 25, 6020 Innsbruck,
	Austria}
\author{Tracy E. Northup}
\address{Institut f{\"u}r Experimentalphysik, Universit{\"a}t Innsbruck, Technikerstra\ss e 25, 6020 Innsbruck,
	Austria}

\date{\today}

\begin{abstract}
We confine a microparticle in a hybrid potential created by a Paul trap and a dual-beam optical trap. We transfer the particle between the Paul trap and the optical trap at different pressures and study the influence of feedback cooling on the transfer process. This technique provides a path for experiments with optically levitated particles in ultra-high vacuum and in potentials with complex structure.
\end{abstract}

\maketitle
\section{Introduction}
Microparticles and nanoparticles levitated in vacuum are promising experimental platforms for testing fundamental physics and building sensitive detectors~\cite{gonzalezballestero2021levitodynamics,millen2020optomechanics}. 
Levitation can be realized with optical~\cite{ashkin1970acceleration}, electric~\cite{paul1990electromagnetic}, or magnetic forces~\cite{hsu2016cooling}, each of which has benefits and drawbacks~\cite{gonzalezballestero2021levitodynamics}.
Optical trapping provides strong confinement; however, the trapping region is typically limited to a few cubic micrometers. Paul traps provide deep and wide potentials at the cost of low trapping stiffness. Magnetic traps do not exploit any oscillating fields, which could be beneficial for the coherence of the particles' motion, but resonance frequencies in such traps are typically below \SI{1}{\kilo\hertz}. Combining different traps provides the possibility to exploit the benefits of each technique while avoiding the drawbacks. For example, the optical field of a high-finesse cavity has been used to trap a particle and cool its motion while the deep and wide potential of a Paul trap acted as a safety net if the particle was lost from the optical trap~\cite{millen2015cavity}. A ``dimple" trap has also been created that combined tight particle confinement with reduced bulk heating by bringing together optical tweezers and a Paul trap~\cite{conangla2020extending}.
For atomic ions, hybrid electro-optical traps, first demonstrated more than a decade ago, have opened up prospects for ultracold chemistry studies and for quantum simulations using tailored potentials~\cite{schaetz2017trapping}.

Here we demonstrate a hybrid electro-optical trap for microparticles in which a dual-beam optical trap is superimposed on a Paul trap. The traps can operate simultaneously, or the potential of one trap can be switched off while the potential of the other is kept on. A levitated silica microsphere is transferred back and forth between the two traps. 
Our demonstration is in low vacuum but could be extended to mesoscopic particles trapped in ultra-high vacuum (UHV)~\cite{bykov2019direct}. Hybrid traps can also be combined with dynamic shaping of confining potentials, as proposed 
for large delocalizations of levitated particles~\cite{weiss2021large}.

\section{Experimental setup}
\begin{figure}
	\centering
	\includegraphics[width=0.95\linewidth]{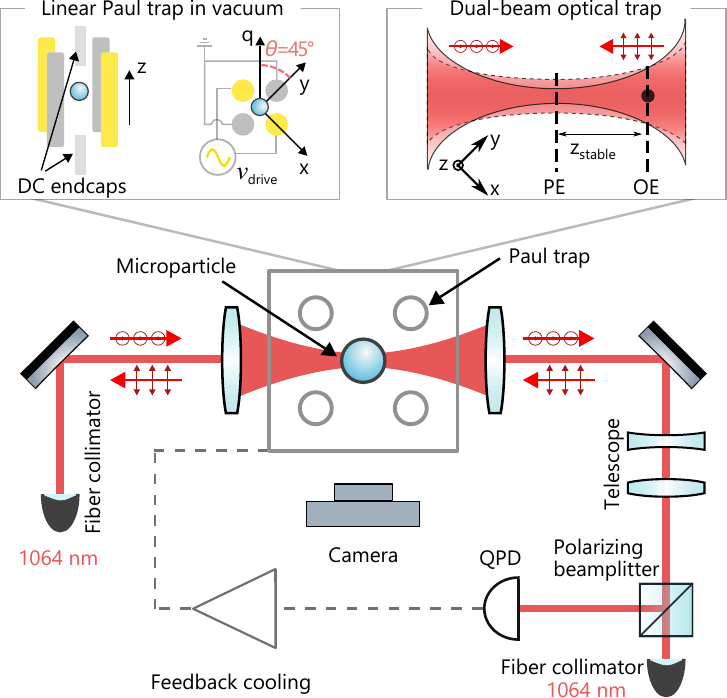}
	\caption{Schematic of the experimental setup. Two counter-propagating laser beams with orthogonal polarizations form an optical trap for a microparticle. The optical trap is superimposed on a linear Paul trap. One beam of the dual-beam optical trap is reflected onto a quadrant photodiode (QPD) in order to detect the particle's center-of-mass motion. Additionally, the particle can be imaged with a camera. Electrodes mounted next to the Paul trap are used as force actuators to apply feedback cooling. The insets show the orientations of the two traps with respect to the lab frame of reference. In the right inset, the particle's equilibrium positions in the Paul trap and the optical trap are indicated with PE and OE respectively.
	}
	\label{fig:fig_1}
\end{figure}
A schematic overview of the experimental setup is shown in Fig.~\ref{fig:fig_1}. The linear Paul trap is mounted in a vacuum chamber; the distance between opposite radiofrequency (RF) electrodes is $2r_0 = \SI{3}{\milli\meter}$, while the distance between endcap electrodes is $2z_0 = \SI{7.9}{\milli\meter}$. The trap is driven with a peak-to-peak voltage of \SI{600}{\volt} at \SI{2}{\kilo\hertz}, with  \SI{10}{\volt} on the endcap electrodes. Two \SI{1064}{\nano\meter} laser beams with orthogonal polarization are delivered to the experimental setup via polarization-maintaining optical fibers. 
The vertically polarized laser beam is focused on the particle through the ion-trap electrodes with a lens of \SI{75}{\milli\meter} focal length, corresponding to a numerical aperture (NA) of 0.25. The horizontally polarized beam is reduced with a telescope and focused on the particle through the opposite pair of electrodes with a lens of the same focal length, corresponding to an NA of 0.05. Counter-propagating beams with different NAs form a stable optical trap~\cite{schmidt2012metrology}. The particle position is monitored with a camera; the detection plane is defined by the vectors $\hat{z}$ and $\hat{q}$ illustrated in Fig.~\ref{fig:fig_1}. Light from the vertically polarized beam is guided by a set of mirrors and a polarizing beam splitter to a quadrant photodiode (QPD), providing interferometric detection of particle position~\cite{gittes1998interference}. We have the option to cool the particle via a feedback voltage that is derived from the QPD position signal and applied to electrodes mounted next to the Paul trap~\cite{dania2021optical}. 
\section{Results and discussion}
\begin{figure}[t]
	\centering
	\includegraphics[width=1\linewidth]{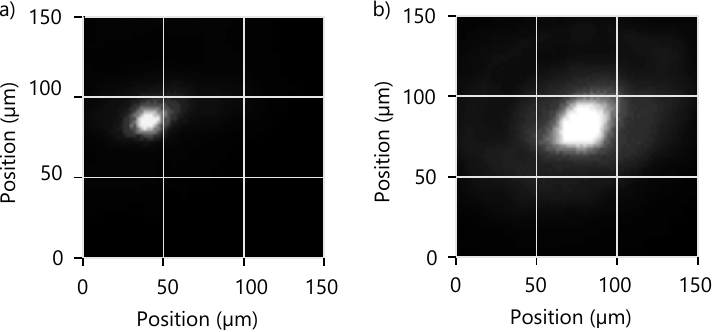}
	\caption{A microparticle trapped (a) in the Paul trap and (b) in the optical dual-beam trap. The distance between the particle's positions is \SI{38(2)}{\micro\meter}.}
	\label{fig:fig_2}
\end{figure}

Having summarized the experimental setup, we now describe the procedure for particle transfer between the Paul trap and the optical trap. We study the transfer process in two pressure regimes: below and above \SI{1}{\milli\bar}, which we refer to as low and high pressure. We start in the low-pressure regime, where we load a silica microparticle \SI{3\pm0.2}{\micro\meter} in diameter\footnote{Bangs Laboratories Inc.} into the Paul trap via laser-induced acoustic desorption~\cite{AsenbaumKuhnNimmrichterEtAl2013,Millen2016,bykov2019direct}. The particles typically carry about 5000 elementary charges after loading. At a pressure of \SI{5e-2}{\milli\bar}, residual gas provides enough damping to slow down the particles within the trap volume; the temporal control of the Paul-trap potential described in Ref.~[\onlinecite{bykov2019direct}] is not needed. At the same time, the damping is sufficiently small that desorbed particles still reach the trap. 
 
After loading, we leak air into the vacuum chamber until a pressure of  \SI{14}{\milli\bar} is reached, thus entering the high-pressure regime. We align the vertically polarized beam to the particle by adjusting the 3D~translational stage on which the focusing lens is mounted while maximizing the light scattered by the particle and imaged on the camera. An image of the particle trapped in the Paul-trap potential is shown in Fig.~\ref{fig:fig_2}a. At this point, we use low optical power, on the order of a few milliwatts in each arm. Thus, the laser beams produce negligible optical forces, and the light is solely for particle detection. 
To align the second beam's path to that of the first, we maximize the coupling of light from the second beam into the output collimator of the first, adjusting only components that do not affect the alignment of the first beam. At this point, the two beam paths overlap and the beam waists are aligned to the particle at the center of the Paul trap.

To form a stable optical trap, we increase the power of the \SI{1064}{\nano\meter} laser to between \SI{100}{\milli\watt} and \SI{200}{\milli\watt} in each beam. An image of an optically trapped particle is shown in Fig.~\ref{fig:fig_2}b. As illustrated in the right inset of Fig.~\ref{fig:fig_1}, the potential minima of the Paul trap and the optical trap do not coincide. As a result, optical trapping displaces the particle from the Paul trap center. The difference of the particle positions in Fig.~\ref{fig:fig_2}a and Fig.~\ref{fig:fig_2}b is $z_{\text{stable}}=\SI{38(2)}{\micro\meter}$, in good agreement with the expected displacement of \SI{30(6)}{\micro\meter} calculated from the NA values. 
\begin{figure}[t]
	\centering
	\includegraphics[width=1\linewidth]{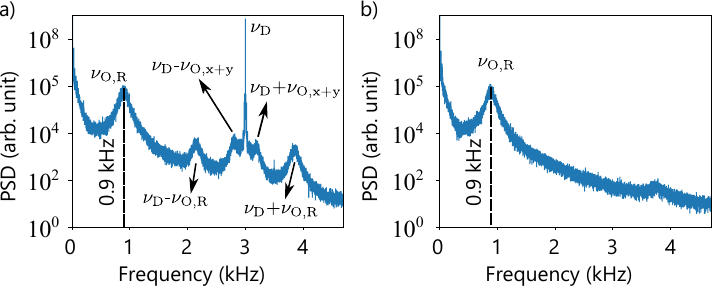}
	\caption{Power spectral densities of the microparticle trapped in the optical trap when the Paul trap is (a) active and (b) inactive.  The Paul-trap drive frequency and optical trap frequency are indicated by $\nu_{\text{D}}$ and $\nu_{\text{O}}$ respectively for both $x+y$ and radial particle motion.}
	\label{fig:fig_3}
\end{figure}

Next, we analyze the motion of the particle in the optical trap. Fig.~\ref{fig:fig_3}a shows the power spectral density (PSD) of the particle's center-of-mass (CoM) motion in the optical trap while the Paul trap is active. The sharp peak at frequency $\nu_{\text{D}} = \SI{3}{\kilo\hertz}$ corresponds to the Paul-trap drive frequency. The resonance of the particle's motion in the radial plane of the optical trap, i.e., in the plane defined by the vectors $\hat{z}$ and $\hat{x} - \hat{y}$, is at frequency $\nu_\text{{O,R}} = \SI{0.9}{\kilo\hertz}$. The $z$ and $x-y$ trap frequencies are degenerate due to the symmetry of the dual-beam trap along the beam propagation axis. The radial motion produces sidebands around $\nu_{\text{D}}$ at frequencies \SI{2.1}{\kilo\hertz} and \SI{3.9}{\kilo\hertz}. From the radial frequencies and the beam diameter of \SI{14}{\micro\meter} at the particle's equilibrium position, we estimate the potential depth to be \SI{30}{\electronvolt}. Particle oscillations in the beam propagation direction $\hat{x}+\hat{y}$ are at frequency $\SI{0.2}{\kilo\hertz}$. The peak of the motion along this direction is beneath the low-frequency noise of the detection system. However, sidebands produced by this motion are visible at frequencies $\SI{2.8}{\kilo\hertz}$ and $\SI{3.2}{\kilo\hertz}$. We cannot assign a potential to the forces acting along the beam propagation direction because they are non-conservative scattering forces~\cite{novotny2012principles}, but we can estimate the work needed to remove the particle from the trap: \SI{20}{\electronvolt}.

As the last step of the particle transfer at high pressure, we switch off the Paul trap drive by ramping down the voltage linearly over \SI{5}{\second}. We keep the endcap voltage on since the field of the endcap electrodes has a negligible influence on the particle in the optical trap. Figure~\ref{fig:fig_3}b shows the PSD of the particle motion while the Paul trap is inactive. Only a peak corresponding to radial motion in the optical trap is present. The peak coincides with the peak in Fig.~\ref{fig:fig_3}a, which suggests that the influence of the Paul trap on the optically trapped particles is negligible. The small bump around \SI{4}{\kilo\hertz} is produced by the noise of the detection system. We can also transfer the particle back from the optical trap to the Paul trap by switching on the Paul-trap potential and decreasing the optical power in the dual-beam trap.
\begin{figure}[t]
	\centering
	\includegraphics[width=1\linewidth]{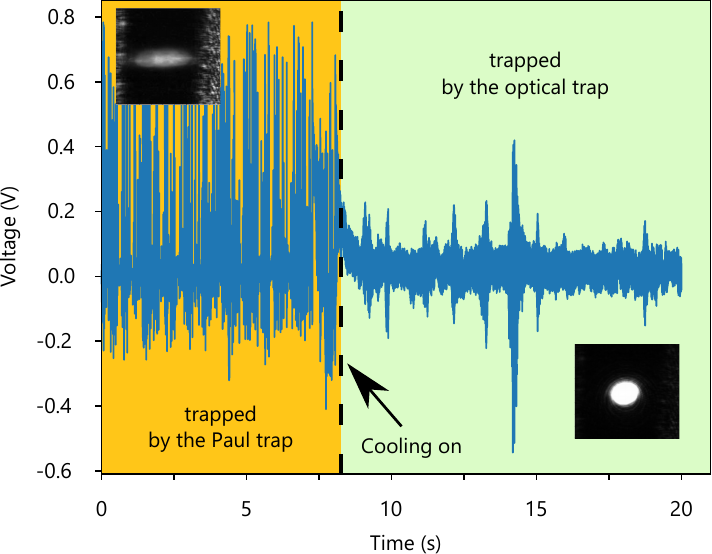}
	\caption{Time trace of the particle's motion at \SI{5e-2}{\milli\bar} measured with the QPD, in the presence of both the Paul trap and the optical trap.  Feedback cooling is turned on at the time indicated by the dashed line. Insets: camera images of the particle's motion.
		\label{fig:fig_4}}
\end{figure}

Finally, we study the transfer process at low pressure. Below $\SI{1}{\milli\bar}$, the motion of the particle in the optical trap becomes unstable,
as also observed in other experiments~\cite{kiesel2013cavity,millen2014nanoscale,price2015invacuo}. Feedback cooling allows the particle to be stabilized at these pressures~\cite{li2011millikelvin,gieseler2012subkelvin}. Here, we apply electrical feedback cooling along the $\hat{x} - \hat{y}$ direction~\cite{dania2021optical}, which allows us to transfer the particle from the Paul trap to the optical trap at \SI{5e-2}{\milli\bar}. Transfer at lower pressures might be possible if we were to apply feedback cooling along all three spatial directions. However, the current experimental setup does not allow it: electronics to extract the motion along the beam propagation direction are not installed, and the feedback force lies in the plane defined by the vectors $\hat{z}$ and $\hat{x} - \hat{y}$. Figure~\ref{fig:fig_4} shows a time trace measured with the QPD during the transfer process. The region on the left side of the dashed line corresponds to the motion of the particle in the Paul trap. The motion is driven by the scattering force produced by the laser beams. The upper-left inset shows a snapshot of this motion captured with the camera. When feedback cooling is activated, the particle is captured by the optical trap. It is then confined to a smaller spatial region, as can be seen from the time trace to the right of the dashed line and from the snapshot in the lower-right inset. In order to provide a safety net in case the particle escapes the optical trap, we keep the Paul trap active during this measurement.

\section{Conclusion}
In conclusion, we have built a hybrid trap for microparticles, combining a linear Paul trap and a dual-beam optical trap. If higher optical powers are used, the method can be extended to particles of smaller size.
A particle was transferred from the Paul trap to the optical trap at pressures above \SI{1}{\milli\bar}, where optical trapping is stable. With the help of feedback cooling along one axis, we also transferred the particle from the Paul trap to the optical trap at \SI{5e-2}{\milli\bar}. We have thus demonstrated a method of loading particles into optical traps that --- when combined with feedback cooling along three axes --- is expected to be compatible with UHV pressures, at which isolation from the environment is sufficient for future experiments in the quantum regime\cite{gonzalezballestero2021levitodynamics,millen2020optomechanics}.

\begin{acknowledgments} 
We thank Alexander Eberhardter and Marius Trojer for helpful discussions. This work was supported by Austrian Science Fund (FWF) Project No. Y951 and by the ESQ Discovery grant “Sympathetic detection and cooling of nanoparticles levitated in a Paul trap” of the Austrian Academy of Sciences.
\end{acknowledgments}

\section*{Author declarations}
\subsection*{Conflict of Interest}
The authors have no conflicts to disclose.
\section*{Data availability}
The data that support the findings of this study are available from the corresponding author upon reasonable request.

\bibliography{biblio_hybrid}
\end{document}